\documentclass[twocolumn,epjc3]{svjour3}          

\RequirePackage[T1]{fontenc}

\smartqed  

\RequirePackage{graphicx}
\RequirePackage{mathptmx}      
\RequirePackage{flushend}
\RequirePackage[numbers,sort&compress]{natbib}
\RequirePackage[colorlinks,citecolor=blue,urlcolor=blue,linkcolor=blue]{hyperref}
\usepackage{amssymb,amsmath,array}
\RequirePackage{xspace}
\usepackage{hyperref}

\mathchardef\Upsilon="7107
\def\Y#1S{\ensuremath{\Upsilon{(#1S)}}\xspace}

\def\Dbar    {\kern 0.2em\overline{\kern -0.2em D}{}\xspace}

\def\Bbar    {\kern 0.18em\overline{\kern -0.18em B}{}\xspace}

\def\epem {\ensuremath{e^+e^-}\xspace}

\newcommand{\kev}{\ensuremath{\mathrm{\,ke\kern -0.1em V}}\xspace}
\newcommand{\mev}{\ensuremath{\mathrm{\,Me\kern -0.1em V}}\xspace}
\newcommand{\mevcc}{\ensuremath{{\mathrm{\,Me\kern -0.1em V\!/}c^2}}\xspace}
\newcommand{\gev}{\ensuremath{\mathrm{\,Ge\kern -0.1em V}}\xspace}
\newcommand{\gevcc}{\ensuremath{{\mathrm{\,Ge\kern -0.1em V\!/}c^2}}\xspace}
\newcommand{\tev}{\ensuremath{\mathrm{\,Te\kern -0.1em V}}\xspace}
\newcommand{\tevcc}{\ensuremath{{\mathrm{\,Te\kern -0.1em V\!/}c^2}}\xspace}
\newcommand{\ev}{\ensuremath{\mathrm{\,e\kern -0.1em V}}\xspace}

\newcommand{\EP}{\ensuremath{e^+}\xspace}
\newcommand{\EM}{\ensuremath{e^-}\xspace}
\newcommand{\EPEM}{\ensuremath{e^+e^-}\xspace}
\newcommand{\EMEM}{\ensuremath{e^-e^-}\xspace}
\newcommand{\GG}{\ensuremath{\gamma\gamma}\xspace}
\newcommand{\gggg}{\ensuremath{\gamma\gamma \to \gamma\gamma}\xspace}

\newcommand{\LEE}{\ensuremath{\mathcal{L}_{ee}}\xspace}

\newcommand{\WGG}{\ensuremath{W_{\gamma\gamma}}\xspace}

\newcommand{\be}{\begin{equation}}
\newcommand{\ee}{\end{equation}}
\newcommand{\bc}{\begin{center}}
\newcommand{\ec}{\end{center}}
\newcommand{\bi}{\begin{itemize}}
\newcommand{\ei}{\end{itemize}}
\newcommand{\ben}{\begin{enumerate}}
\newcommand{\een}{\end{enumerate}}

\journalname{Eur. Phys. J. C}

\begin{document}

\title{\boldmath Impact of QED backgrounds on light-by-light scattering measurements at $e^+e^-$ and $e^-e^-$ colliders.
}


\author{K.~I.~Beloborodov\thanksref{addr1,addr2}
        \and
        T.~A.~Kharlamova\thanksref{addr1,addr2}
        \and
        V.~I.~Telnov~\thanksref{addr1,addr2,c1}
}

\thankstext[$\star$]{c1}{Corresponding author, e-mail:telnov@inp.nsk.su}

\institute{Budker Institute of Nuclear Physics, 630090, Novosibirsk, Russia\label{addr1}
          \and
          Novosibirsk State University, 630090, Novosibirsk, Russia\label{addr2}
}


\date{Received: date / Accepted: date}

\maketitle

\begin{abstract}
The process of elastic scattering of photons ($\gamma\gamma \to \gamma\gamma$, light-by-light scattering) has attracted significant interest in recent years as a loop-induced process sensitive to all charged particles. To date, this process has been studied through Delbr\"{u}ck scattering, photon splitting in the nuclear Coulomb field, and ultra-peripheral heavy-ion collisions at the LHC. Future precision measurements are anticipated at high-luminosity  $e^+e^-$ colliders (SuperKEKb, FCC, CEPC, ILC, CLIC), their $e^-e^-$ options, and $\gamma\gamma$ colliders based on the backscattering of laser photons. In this paper, we show that background QED processes severely limit the study of elastic light-by-light scattering at $e^+e^-$ colliders. For the $e^+e^-$ case, the main background arises from $e^+e^-$ annihilation into two photons within the detector acceptance after the initial electron and positron emit hard ISR photons at small angles. Since only two photons with a small total transverse momentum are registered in the detector in both cases, this process effectively mimics the $\gamma\gamma \to \gamma\gamma$ signal with a significantly larger cross section at high invariant masses. Other relevant background QED processes for both types of collisions are also comprehensively analyzed and discussed.
\end{abstract}

\section{Introduction}

The scattering of light-by-light in a vacuum is one of the most fundamental processes in quantum electrodynamics. Its cross section was theoretically predicted more than half a century ago~\cite{Euler,Heisenberg,Akhiezer,DeTollis,berestetskii}. While the cross section is extremely small for optical photons, it becomes accessible for measurements at photon energies exceeding $m_e c^2$.

\begin{figure}[htbp]
\centering
\vspace{-6mm}
\includegraphics[width=5.5cm]{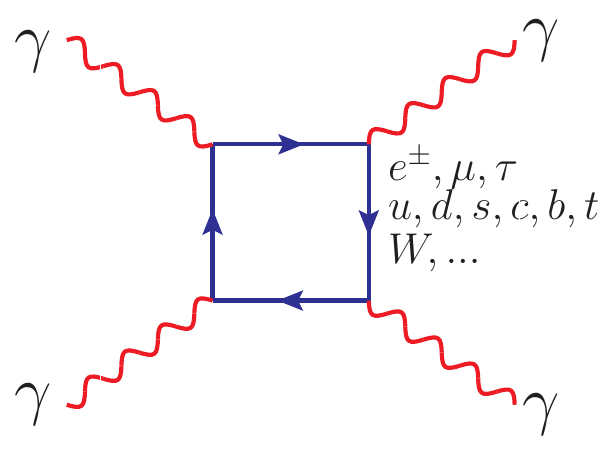}
\caption{Diagram for light-by-light scattering.}
\vspace{-0.3cm}
\label{box}
\end{figure}

\begin{figure}[htbp]
\centering
\includegraphics[width=0.49\textwidth]{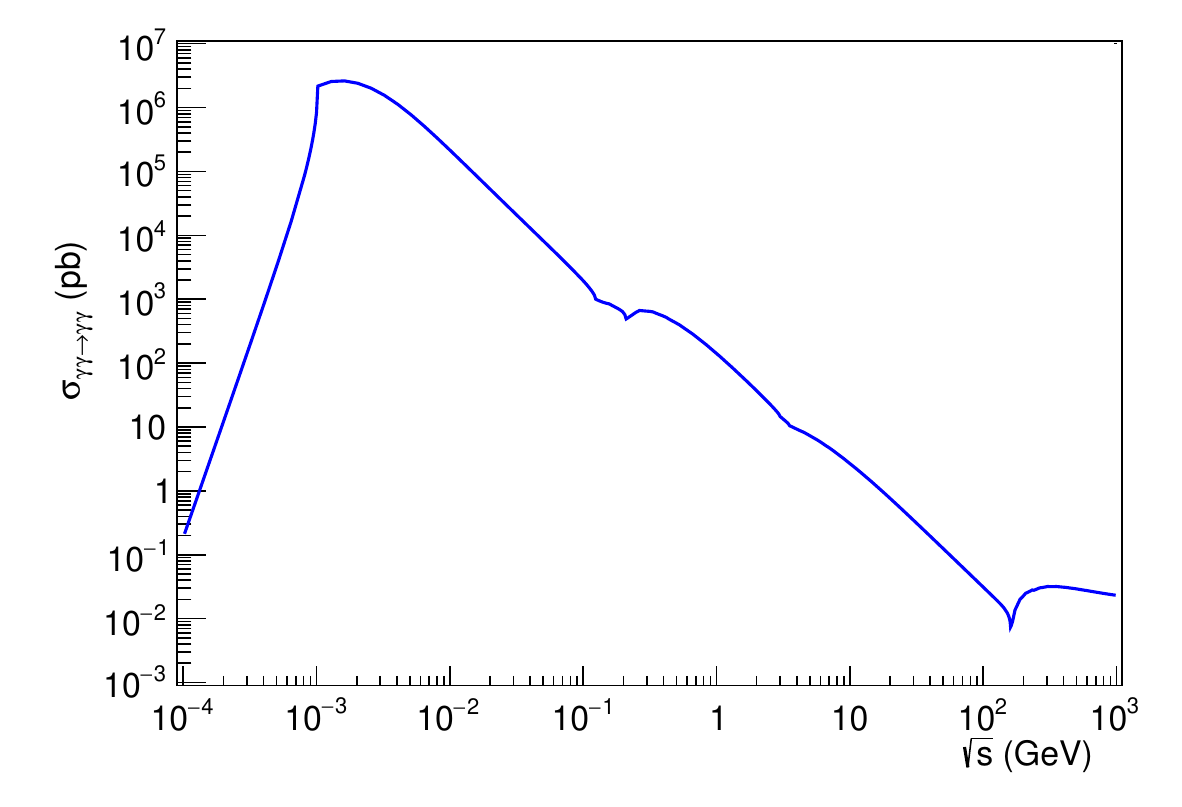} \vspace{-2mm}
\includegraphics[width=0.49\textwidth]{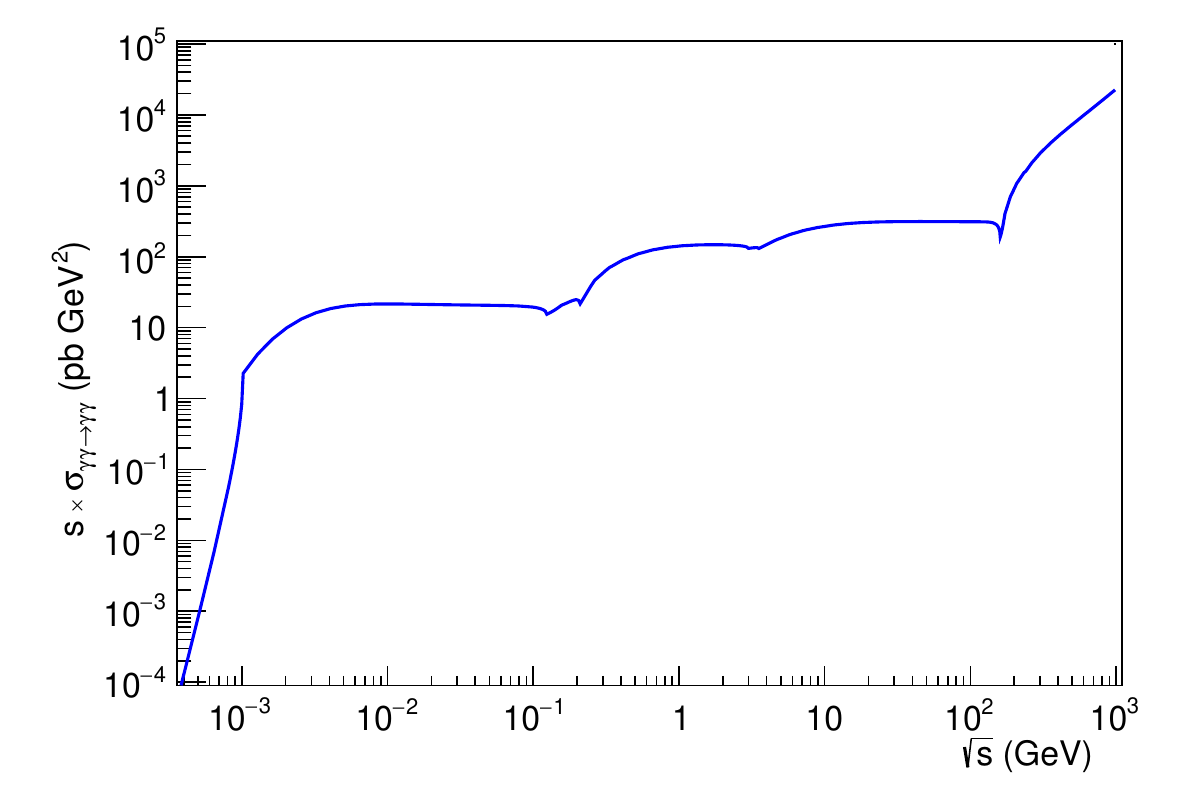}
\caption{Top: the cross section for light-by-light scattering as a function of $\sqrt{s}=\WGG$, bottom: the same cross section multiplied by s.}
\vspace{-0.2cm}
\label{ggggsig}
\end{figure}

The elastic scattering of photons proceeds via the box diagram shown in Fig.~\ref{box}, where all charged particles contribute to the amplitude proportional to the fourth power of their electric charge. The cross section of this process is shown in Fig.~\ref{ggggsig} (top) and Fig.~\ref{ggggsig} (bottom) multiplied by the squared center-of-mass energy $s=\WGG^2$; the plots clearly demonstrate the threshold effects from various particles. At low energies, quarks are not yet asymptotic; for simplicity, we have included $u$ and $d$ quarks at the $\pi$-meson mass scale and the $s$-quark at the $K$-meson mass scale. Since the amplitudes from different fermions interfere, the total cross section is proportional to $(\sum e_i^2)^2$. At energies $\WGG > 2m_W c^2$, the dominant contribution arises from $W$-boson loops.

As shown, the \gggg scattering cross section is significant at energies of 1--100 MeV. It was proposed half a century ago to study this process at \EPEM storage rings through the collision of quasi-real virtual photons~\cite{Baier1975}. Although two-photon physics at \EPEM colliders has been actively studied since the 1970s, no systematic attempts were made to observe elastic photon-photon scattering via the box diagram; processes of the type $\gamma^*\gamma^* \to \gamma\gamma$ were observed only through $C$-even resonance production.

Related to elastic \gggg scattering are Delbr\"{u}ck scattering (elastic photon scattering by a nucleus) and photon splitting in a nuclear field, which also proceed via the box diagram. These processes were thoroughly investigated at VEPP-4M in Novosibirsk using a tagged photon beam with energies of 120--450 MeV~\cite{Akhmadaliev-D,Akhmadaliev-PS}.

Renewed interest in the \gggg process is associated with the development of high-energy linear \EPEM colliders and photon colliders~\cite{GKST81-r,GKST81-e,GKST83,TESLATDR,Ginzburg-1982,Jikia,Gounaris,Ginzburg-mon,Bern,Bardin,Telnov-12, Beloborodov,Ellis,Inan}. Furthermore, ultra-peripheral ion-ion collisions at the LHC offer promising opportunities for studying light-by-light scattering~\cite{dEnterria-2013,Klusek,Klusek-low,Lebiedowicz,Ellis-2}.

Following the suggestion in~\cite{dEnterria-2013}, the ATLAS and CMS collaborations performed the first measurements of light-by-light scattering in high-energy heavy-ion collisions at the LHC~\cite{ATLAS-ev,CMS-ev,ATLAS-2,CMS-gggg} at $\WGG \approx$ 5--20 GeV.

There are also proposals to develop dedicated photon colliders~\cite{Micieli, Takahashi, Sangal} to study the \gggg process at $\WGG \approx 1$~MeV, where the cross section reaches its maximum of approximately 2~$\mu$b.

There is interest in studying this process at high-luminosity \EPEM colliders (SuperKEKB, FCC-ee, CEPC) and future linear colliders (ILC, CLIC), as well as dedicated \GG facilities, under the assumption that they provide optimal conditions for measuring \gggg scattering. Recently, D. d'Enterria and H.S. Shao investigated the possibility of observing true tauonium in \GG collisions at SuperKEKB and FCC-ee~\cite{Enterria2022}, where the tauonium is detected via its decay into two photons. In their study, \gggg scattering was considered one of the primary backgrounds alongside $C$-even charmonium resonance decays into photon pairs~\cite{Enterria2022,dEnterria-res}. In the present paper, we re-examine this assumption and investigate whether \gggg scattering is indeed the dominant background process at \EPEM colliders in the energy range $\WGG \sim 3.5$~GeV for events with only two photons within the detector acceptance and small total transverse momentum.

\begin{figure}[htbp]
\centering
\vspace{-5mm}
\includegraphics[width=5cm]{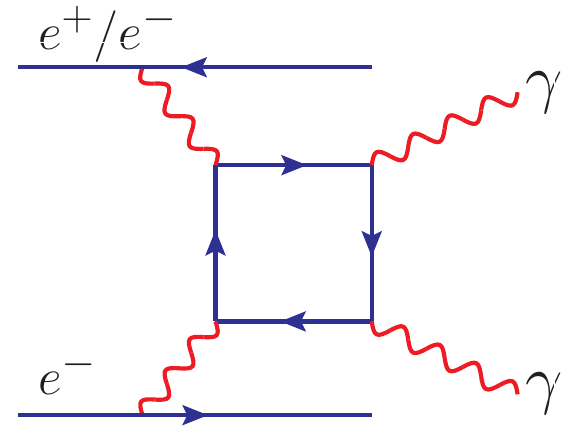}
\caption{Diagram for light-by-light scattering in \epem collisions.}
\vspace{-0.2cm}
\label{ee-box}
\end{figure}

\begin{figure}[htbp]
\centering
\includegraphics[width=5cm]{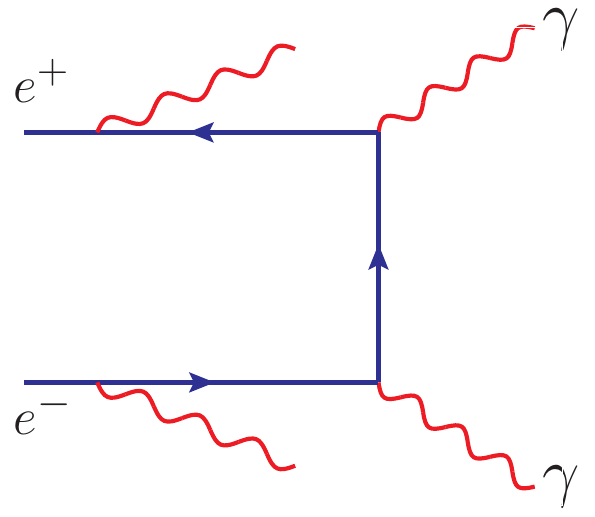}
\caption{\epem annihilation diagram in which photons are completely indistinguishable from the case of light by light scattering.}
\label{ee-gg}
\end{figure}

   Indeed, \gggg scattering at \EPEM colliders proceeds via the diagram shown in Fig.~\ref{ee-box}. In this process, only the final-state photons are detected at large angles, while the electrons escape the detector at small angles. This process involves six vertices; therefore, its cross section is proportional to $\alpha^6$, where $\alpha = e^2 / (\hslash c) \approx 1/137$. In contrast, the direct annihilation process $\EPEM \to \GG$ with two final-state photons is proportional to $\alpha^2$. In the latter case, the photons have energies equal to the beam energy $E_0$, allowing such events to be easily identified and rejected. Similarly, if only one electron or positron emits an Initial State Radiation (ISR) photon at a small angle, the event remains kinematically distinguishable since one of the initial colliding particles retains the energy $E_0$. However, if both the \EP and \EM emit ISR photons, as illustrated in Fig.~\ref{ee-gg}, the resulting process -- with a cross section proportional to $\alpha^4$ -- becomes kinematically indistinguishable from \gggg scattering ($\sigma \propto \alpha^6$). This critical background contribution appears to have been overlooked even in the pioneering study by QED experts~\cite{Baier1975} dedicated to the feasibility of measuring light-by-light scattering.

In this paper, we evaluate these and other QED backgrounds to \gggg scattering at \EPEM, \EMEM, and \GG colliders. Our analysis combines analytical considerations with numerical simulations performed using CompHEP version 4.5.2~\cite{comphep}. We assume that the two photons from the \gggg process are detected within the angular range $30^{\circ} < \theta < 150^{\circ}$, while all other particles escape within $10^{\circ}$ cones around the beam axis.

 The paper is organized as follows. In Section 2, we calculate the cross section for elastic light-by-light scattering (the signal) at $e^+e^-$ and $e^-e^-$ colliders. Sections 3--6 are devoted to the analysis of the primary background QED processes. Section 7 presents a summary plot comparing the signal and background cross sections under specific selection cuts applied to suppress the background. Finally, Conclusions are given in Section 8.

\section{Cross section for light-by-light scattering at \EPEM and \EMEM colliders}

The process is described by the diagram shown in Fig.~\ref{ee-box}. The spectrum of colliding equivalent photons is given by the well-known formula~\cite{Budnev}:
\begin{equation}
dn_{\gamma} \approx \frac{\alpha}{\pi} \left[ \left(1-x+ \frac{x^2}{2}\right) \ln{\frac{q^2_{\max}}{q^2_{\min}}} - 1 + x \right] \frac{dx}{x} \equiv f(x)dx,
\label{equiv}
\end{equation}
where $x=\omega/E_0$ is the energy fraction of the equivalent photon, $E_0$ is the initial beam energy, and $q^2_{\min} = m^2 x^2 / (1-x)$. The maximum momentum transfer $|q_{\max}|$ is determined by the experimental cut on the acoplanarity angle $\Delta\phi = \pi - |\phi_2 - \phi_1|$ between the final-state photons. In our calculations, we assume $\Delta\phi = 0.01$~rad. Given the logarithmic dependence on $q^2$, we approximate $q_{\max} \approx (\WGG/2) \Delta\phi$, where the invariant mass of the colliding photons is $\WGG^2 \approx 4\omega_1 \omega_2$.

The \GG luminosity, normalized to the collider luminosity $\LEE$, is given by:
\begin{equation}
dL_{\GG} = \int f(x)f(y)dxdy = 2zdz\int_{x_{\min}}^{x_{\max}} f(x) f\left(\frac{z^2}{x}\right) \frac{dx}{x},
\label{ggdLdw}
\end{equation}
where $x=\omega_1/E_0$, $y=\omega_2/E_0$, $z = \WGG / (2E_0)$ and $z^2 = xy$. The luminosity distribution over $z$ is typically obtained by integrating over $x$ from $z^2$ to 1. However, to account for the detector acceptance, we restrict the maximum rapidity $|\eta| = \frac{1}{2} \ln(x/y)$, which leads to the integration limits $x_{\min} = z e^{-\eta}$ and $x_{\max} = z e^{\eta}$. This rapidity $\eta$ is related to the polar angle $\theta$ as $\eta = -\ln[\tan(\theta/2)]$. For instance, the angle $\theta = 30^{\circ}$ corresponds to $|\eta| = 1.31$. Thus, by constraining the range of $\eta$, we effectively impose angular cuts on the final-state photons without the need for a full kinematic simulation of the scattering angles at this stage.

\begin{figure}[htbp]
\centering
\includegraphics[width=0.48\textwidth]{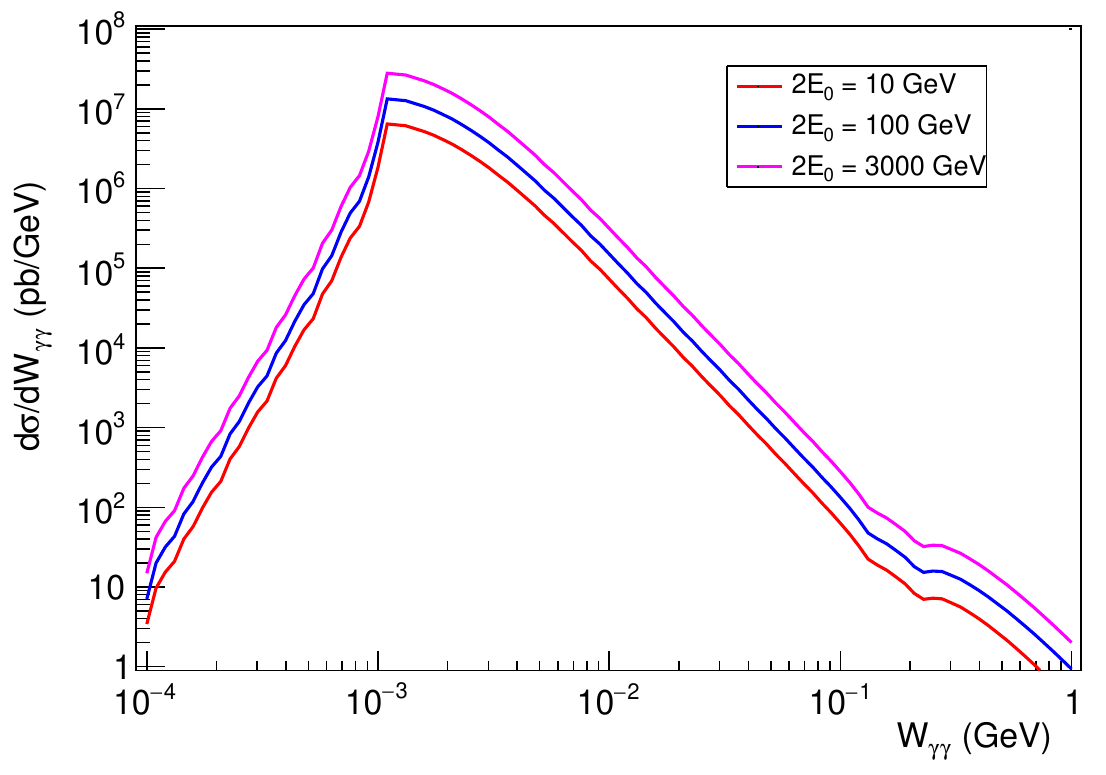}
\includegraphics[width=0.48\textwidth]{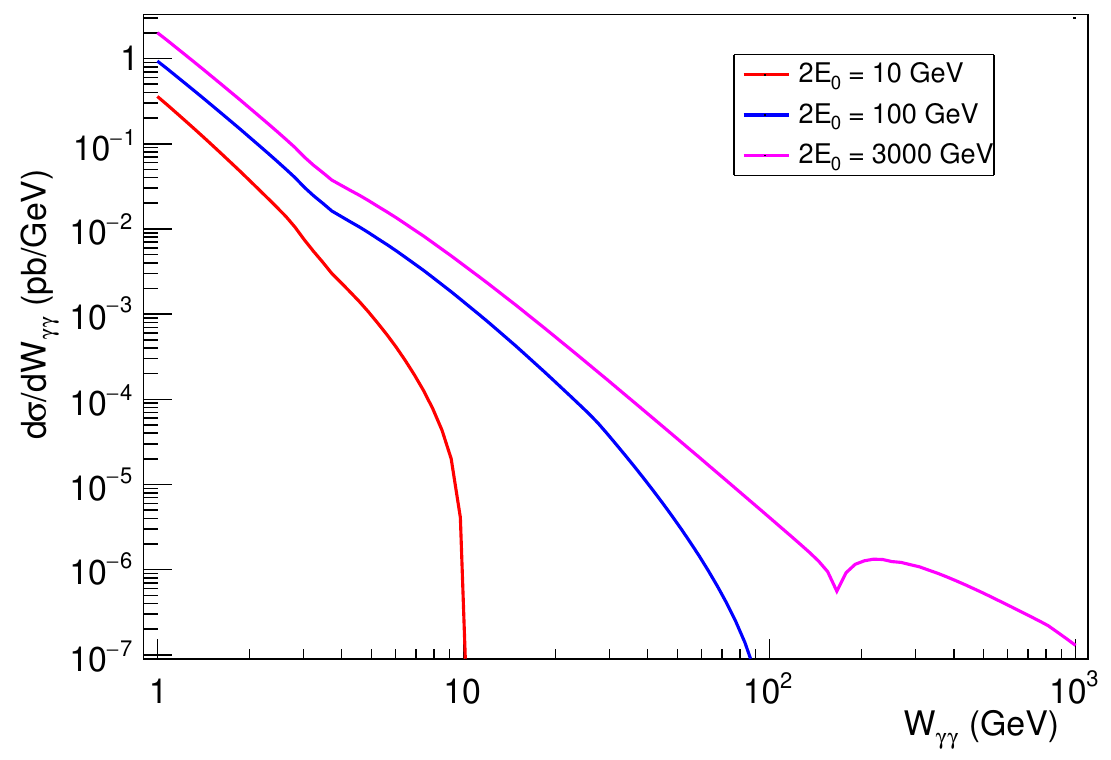}
\caption{The effective cross section for light-by-light scattering in \epem collisions.}
\label{ss-eegg}
\end{figure}

The effective cross section for the $e^+e^- \to e^+e^-\gamma^*\gamma^* \to e^+e^- \gamma\gamma$ process is expressed as:
\be
\frac{d\sigma}{dW} = \frac{\sigma_{\gamma\gamma\to\gamma\gamma}(W)}{2E_0} \frac{dL_{\gamma\gamma}}{dz},
\label{dsdwgg}
\ee
where $\sigma_{\gamma\gamma\to\gamma\gamma}(W)$ is calculated for the angular range $30^{\circ} < \theta < 150^{\circ}$.
The cross section of this process, hereafter referred to as the "signal", is shown in Fig.~\ref{ss-eegg}. In Section 7, it is compared with the distributions of background processes, which are discussed in detail below.

\section{Background process $e^+e^- \to \gamma\gamma\gamma\gamma$} \label{3}

In the process shown in Fig.~\ref{ee-gg}, an electron and a positron emit ISR (initial-state radiation) photons at small angles and then annihilate, producing a pair of photons at large angles. These large-angle photons have small total transverse momentum and a small acoplanarity angle, similar to the photons produced in the $\gamma\gamma \to \gamma\gamma$ process.

The approximate cross section for this process can be derived as follows. The spectrum of electrons/positrons after ISR photon emission is given by the same formula as for equivalent photons (\ref{equiv}), with the substitution $\omega \to E_0 - E$:
\be
dn_e \approx \frac{\alpha}{2\pi} \frac{1+x^2}{1-x} \ln\frac{q^2_{\rm max}}{m^2} dx,
\label{eISR}
\ee
where $x = E/E_0$. The luminosity distribution $dL_{e^+e^-}/dW_{ee}$ can then be obtained in the same manner as in Eq.~(\ref{ggdLdw}).

The cross section for the $e^+e^- \to \gamma\gamma$ process for angles $\theta > \theta_0$ is well known~\cite{berestetskii}:
\be
\sigma_{e^+e^- \to \gamma\gamma} = \frac{2\pi\alpha^2}{W_{ee}^2} \left( \frac{1+\cos \theta_0}{1-\cos \theta_0} - \cos \theta_0 \right).
\ee
For the angular range $30^{\circ} < \theta < 150^{\circ}$, this cross section is:
\be
\sigma_{e^+e^- \to \gamma\gamma} = \frac{2.3 \times 10^{-31}}{W_{ee}^2 \text{ [GeV]}^2} \text{ cm}^2.
\label{se+e-gg}
\ee
Similarly to Eq.~(\ref{dsdwgg}), we have:
\be
d\sigma/dW = \frac{\sigma_{e^+e^- \to \gamma\gamma}(W)}{2E_0} \frac{dL_{e^+e^-}}{dz},
\label{dssdwgg}
\ee
where $z = W_{ee}/2E_0$. Here, $L_{e^+e^-}$ refers to the luminosity for events in which both the electron and positron have emitted ISR photons, normalized to the standard collider luminosity. The cross section of this primary background process, calculated analytically using a rapidity approximation for the detector acceptance, will be presented and compared with the signal in Section 7.

\begin{figure}[htbp]
\centering
\includegraphics[width=4.2cm,height=4.8cm]{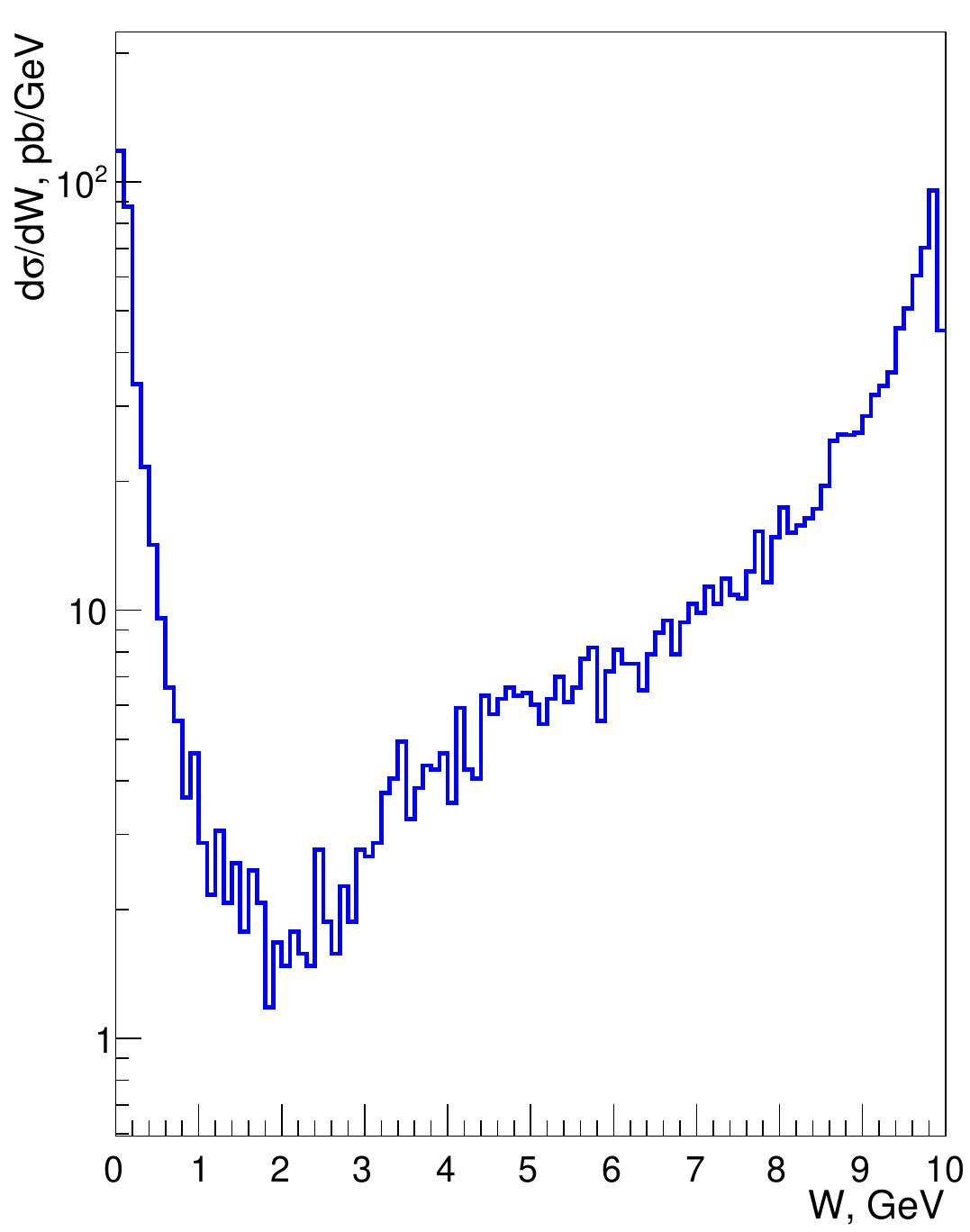} \hspace*{-0.2cm}
\includegraphics[width=4.2cm,height=4.8cm]{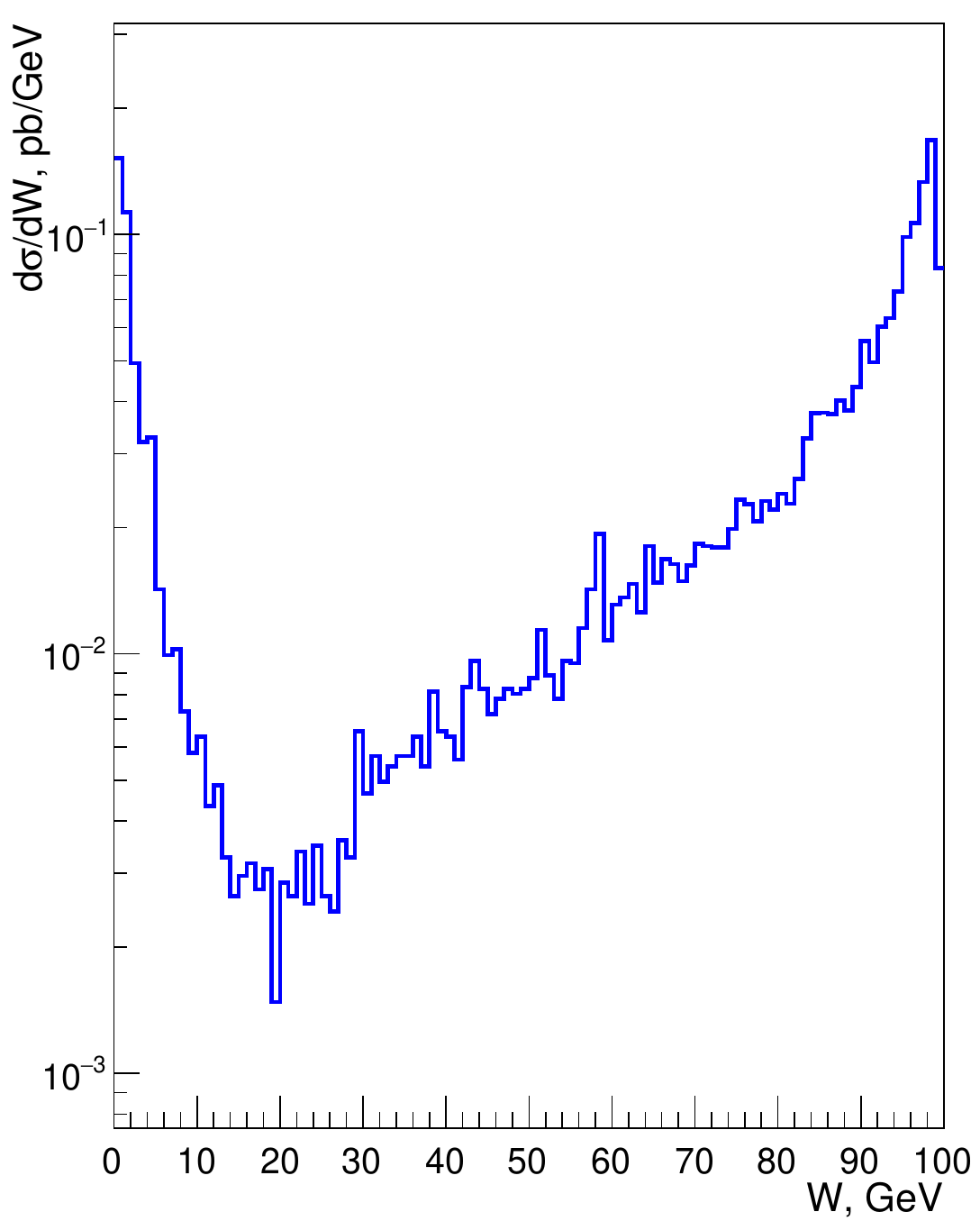}
\caption{Cross section for the $e^+e^- \to \gamma\gamma\gamma\gamma$ process with only angular cuts applied (see text for details).}
\label{dsdw-gggg}
\end{figure}

 To ensure reliability, this important process was also simulated using CompHEP. The following kinematic requirements were applied: two photons must have polar angles in the range $30^{\circ} < \theta < 150^{\circ}$, while the other two must be at small angles, $\theta < 10^{\circ}$ or $\theta > 170^{\circ}$.
Fig.~\ref{dsdw-gggg} shows the cross sections calculated with only these angular cuts applied.

\begin{figure}[htbp]
\centering
\includegraphics[width=4.2cm,height=4.8cm]{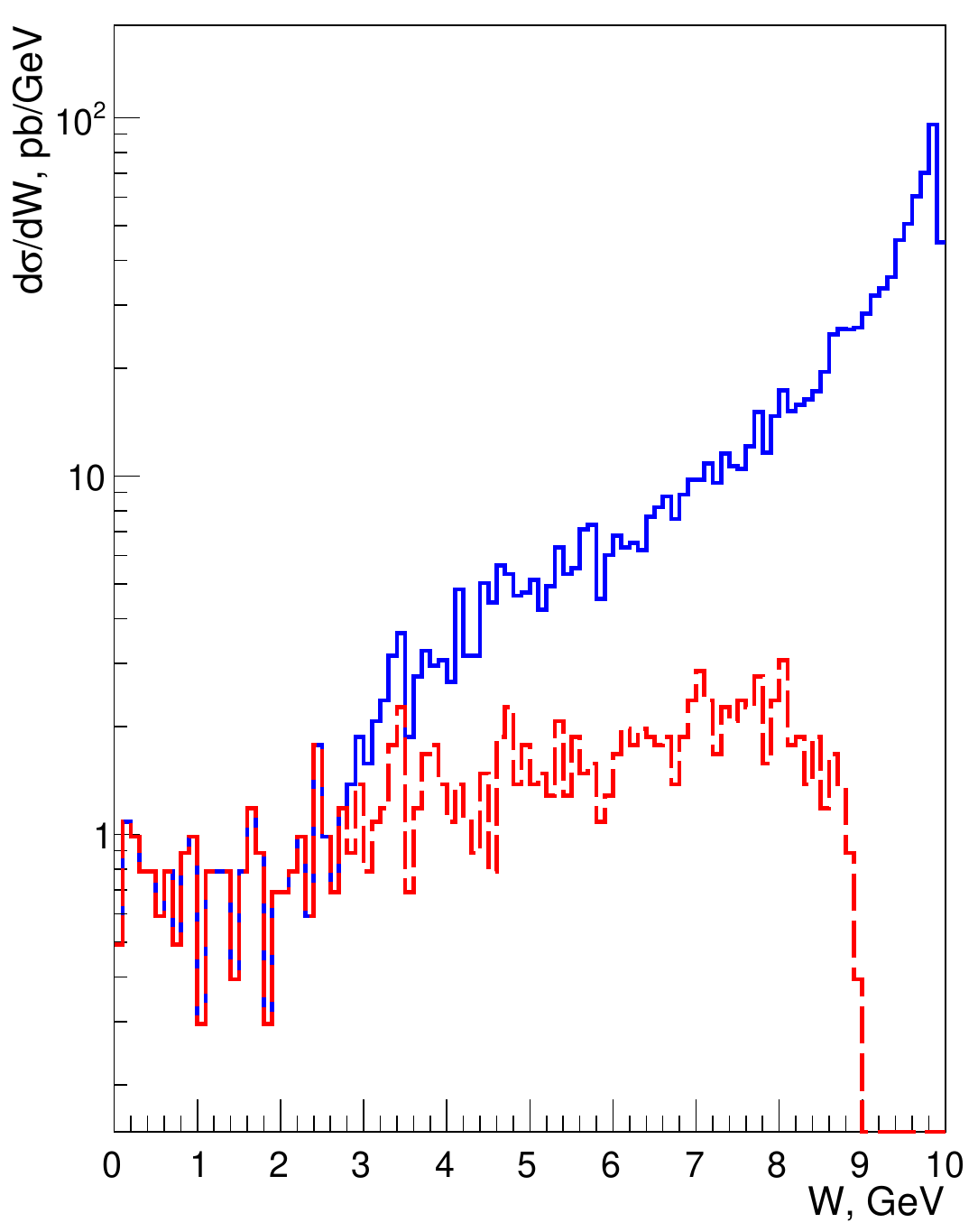} \hspace*{-0.2cm}
\includegraphics[width=4.2cm,height=4.8cm]{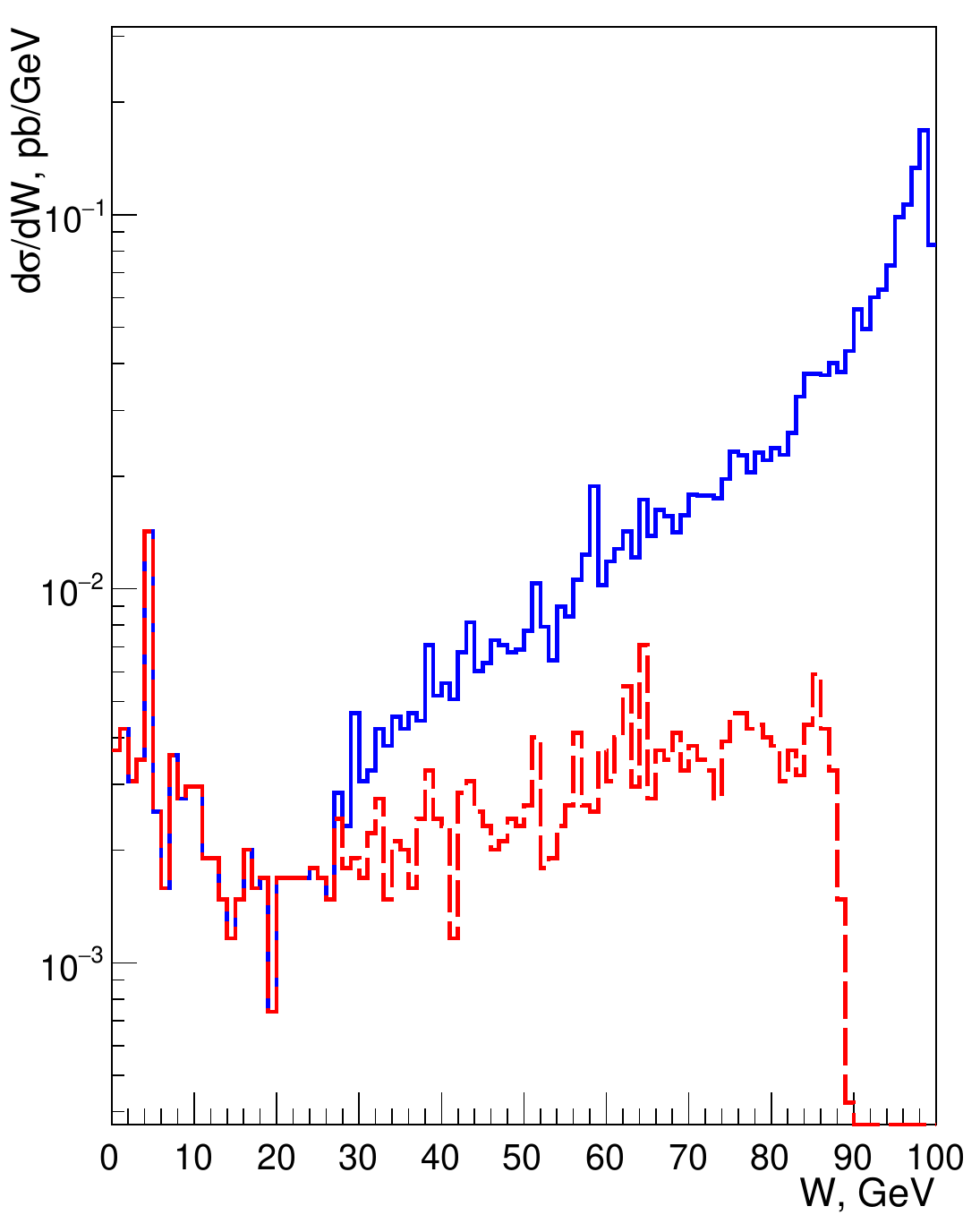}
\caption{Cross section for $e^+e^- \to \gamma\gamma\gamma\gamma$ with the cuts $|p_{t,1}-p_{t,2}|/(p_{t,1}+p_{t,2}) < 0.05$ and $|\Delta \phi| < 0.1$. The red curves include an additional cut $E_e < 0.9E_0$ (see text for details).}
\label{dsdw-gggg-0.05-0.1}
\end{figure}

Fig.~\ref{dsdw-gggg-0.05-0.1} shows the same cross sections as in Fig.~\ref{dsdw-gggg}, but with additional cuts on the relative transverse momentum difference, $|p_{t,1}-p_{t,2}|/(p_{t,1}+p_{t,2}) < 0.05$, and the acoplanarity angle, $|\Delta \phi| < 0.1$. The red curves also include a cut on the maximum energy of the colliding electrons, $E_e < 0.9E_0$, as discussed above.

\begin{figure}[htbp]
\centering
\includegraphics[width=4.2cm,height=4.8cm]{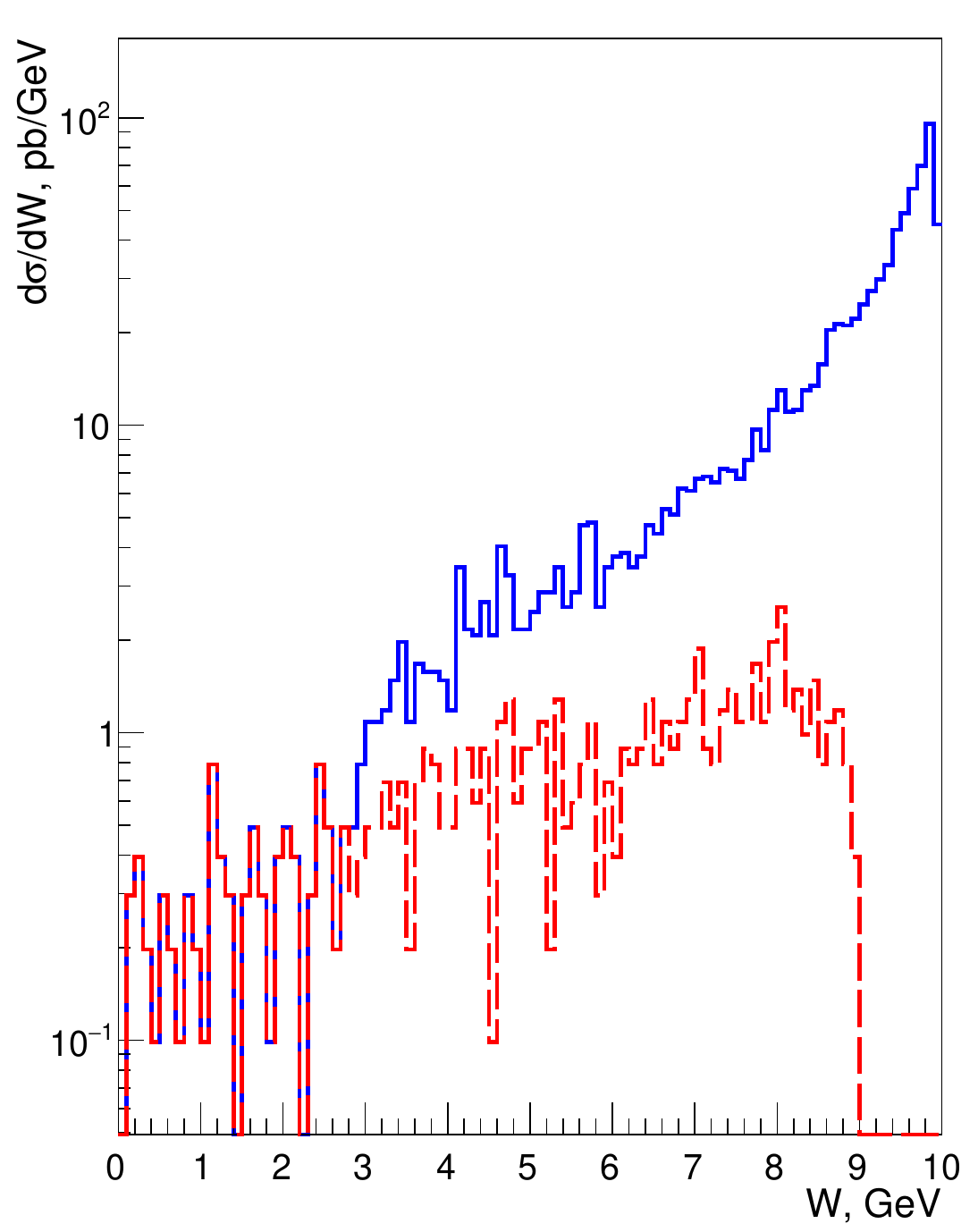} \hspace*{-0.2cm}
\includegraphics[width=4.2cm,height=4.8cm]{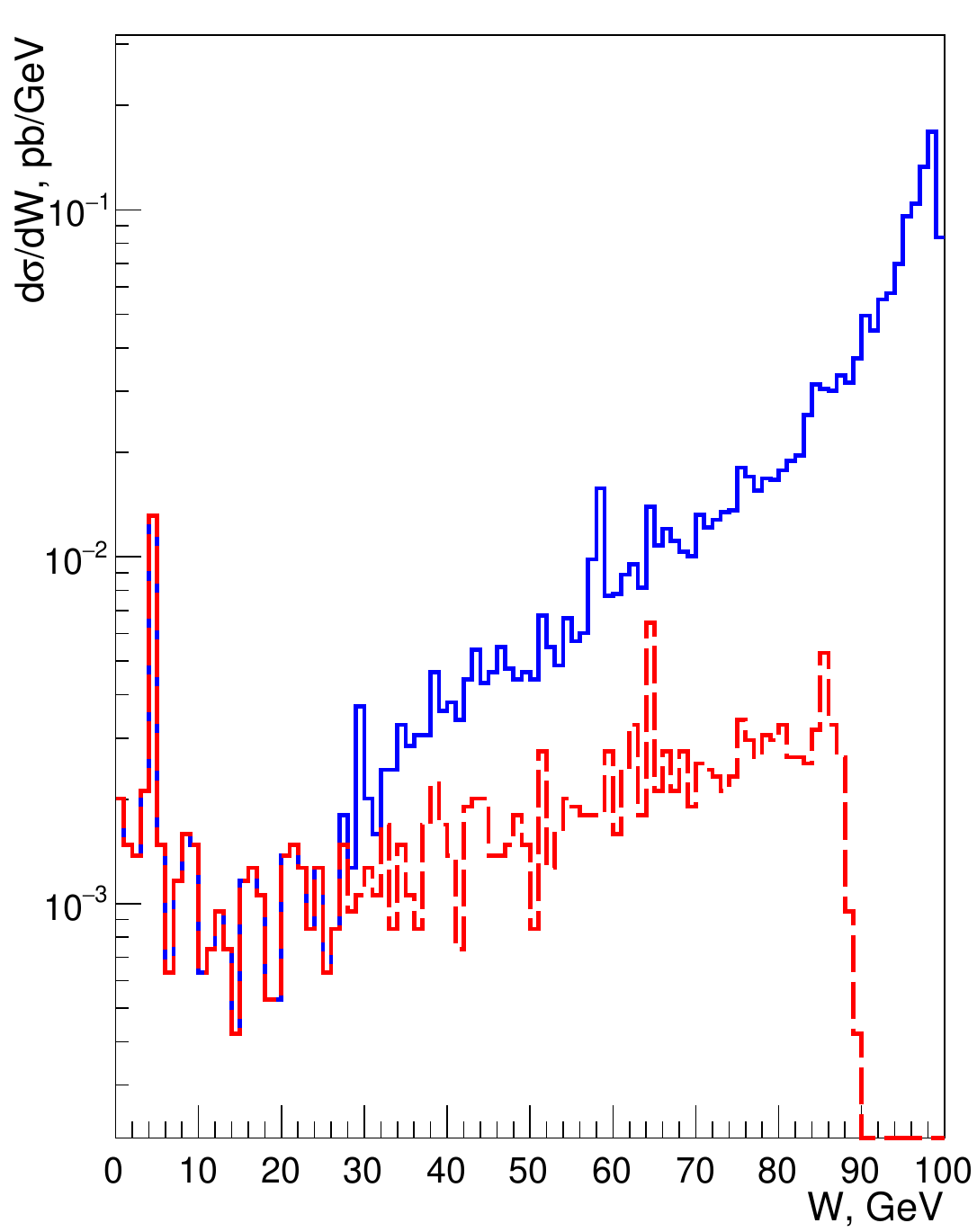}
\caption{Cross section for $e^+e^- \to \gamma\gamma\gamma\gamma$ with the same cuts as in Fig.~\ref{dsdw-gggg-0.05-0.1}, but with a tighter acoplanarity angle cut: $|\Delta \phi| < 0.01$.}
\label{dsdw-gggg-0.05-0.01}
\end{figure}

Fig.~\ref{dsdw-gggg-0.05-0.01} shows the same cross sections as Fig.~\ref{dsdw-gggg-0.05-0.1}, but with a narrower cut on the acoplanarity angle: $|\Delta \phi| < 0.01$. It can be seen that a tighter cut on $|\Delta \phi|$ reduces the cross section at small $W$. The simulation results from CompHEP show good agreement with the analytical calculation, which will be explicitly presented below in Section 7. Some of the observed differences arise from the different treatments of the angular cuts: in CompHEP, the angles are simulated directly, whereas in the analytical calculations, the angular cut was defined in terms of pseudorapidity.

\section{Background process $e^{\pm}e^- \to e^{\pm}e^-\gamma\gamma\gamma$} \label{4}

This process occurs (and is identical) at both $e^+e^-$ and $e^-e^-$ colliders.
\begin{figure}[htbp]
\centering
\includegraphics[width=5cm]{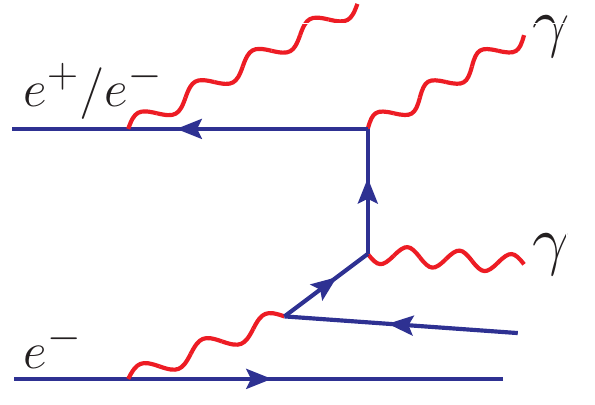}
\caption{Annihilation of an electron/positron that has emitted an ISR photon with a positron/electron from a virtual $e^+e^-$ pair.}
\label{ee-eeggg}
\end{figure}

The most important diagram for this process is shown in Fig.~\ref{ee-eeggg}. Here, an $e^+$ (or $e^-$) that has emitted an ISR photon annihilates with an $e^-$ (or $e^+$) from a virtual $e^+e^-$ pair. This background process is proportional to $\alpha^5$, whereas the signal is $\propto \alpha^6$.

The spectrum of electrons that have emitted an ISR photon was discussed earlier and is given by Eq.~(\ref{eISR}). The distribution of "sea" electrons/positrons within the electron is given by~\cite{Baier-F-K,Fadin-G}:

$$dn_e^e =\left(\frac{\alpha^2}{8\pi^2}\right) \ln^2\left(\frac{q^2_{\rm max}}{m^2}\right) \times$$
\vspace{-5mm}
\be
\qquad \qquad  \times \left[2(1 + x)\ln x+ \frac{4 + 3x - 3x^2 - 4x^3}{3x}\right] dx.
\label{ee--eeggg}
\ee

As before, given the spectra of the colliding $e^+$ and $e^-$, we can calculate the $e^+e^-$ luminosity distribution. By multiplying this distribution by the $e^+e^- \to \gamma\gamma$ cross section from Eq.~(\ref{se+e-gg}), we obtain $d\sigma/dW$. The cross section plot for this background process will be presented and compared with the signal in Section 7.

\section{Background process $e^+e^- \to e^+e^-e^+e^-\gamma\gamma$} \label{5}

\begin{figure}[htbp]
\centering
\includegraphics[width=5cm,height=3.5cm]{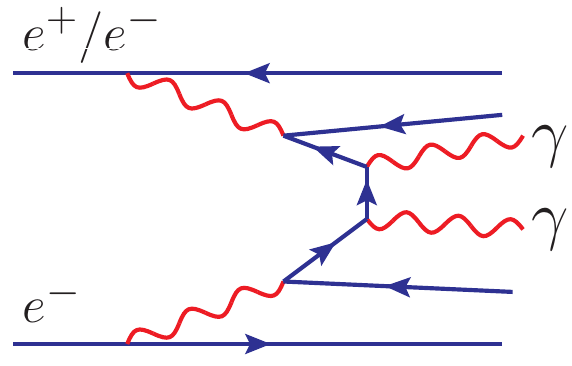}
\caption{Annihilation of virtual (sea) $e^+$ and $e^-$.}
\label{ee-eeeegg}
\end{figure}

The most important diagram for this process is shown in Fig.~\ref{ee-eeeegg}. This process is proportional to $\alpha^6$, the same as the signal. The spectrum of the virtual (sea) $e^{\pm}$ is given by Eq.~(\ref{ee--eeggg}). Proceeding in the same manner as for the other processes, $d\sigma/dW$ can be calculated, and the resulting distribution for this background will be compared with the signal in Section 7.

\section{Background process $e^+e^- \to e^+e^- \gamma\gamma$}\label{6}

\begin{figure}[htbp]
\centering
\includegraphics[width=5cm,height=3.5cm]{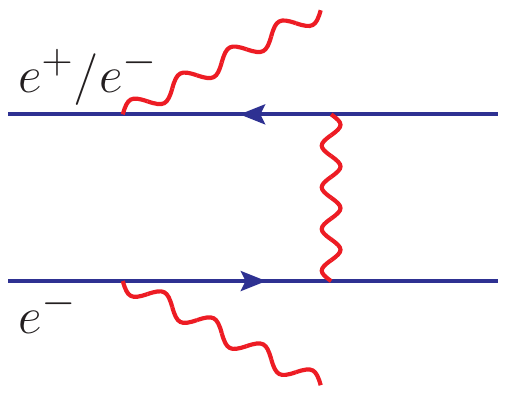}
\caption{Main diagram of the $e^+e^- \to e^+e^- \gamma\gamma$ process.}
\label{dbs}
\end{figure}

This process corresponds to double bremsstrahlung, with the main diagram shown in Fig.~\ref{dbs}; there are approximately 20 other diagrams contributing to the $e^+e^- \to e^+e^- \gamma\gamma$ process. This background is proportional to $\alpha^4$ (whereas the signal is $\propto \alpha^6$) and has the largest total cross section. However, most of the photons are emitted at small angles. Photon pairs emitted at large angles do not typically peak at small total transverse momentum, nor do they exhibit small acoplanarity angles. Therefore, this background can be effectively suppressed or subtracted. While it does not introduce a systematic error to the light-by-light scattering measurement, it may increase the statistical uncertainty. Since analytical estimates for this process are challenging, it has been simulated using CompHEP.

\begin{figure}[htbp]
\centering
\includegraphics[width=4.2cm,height=4.8cm]{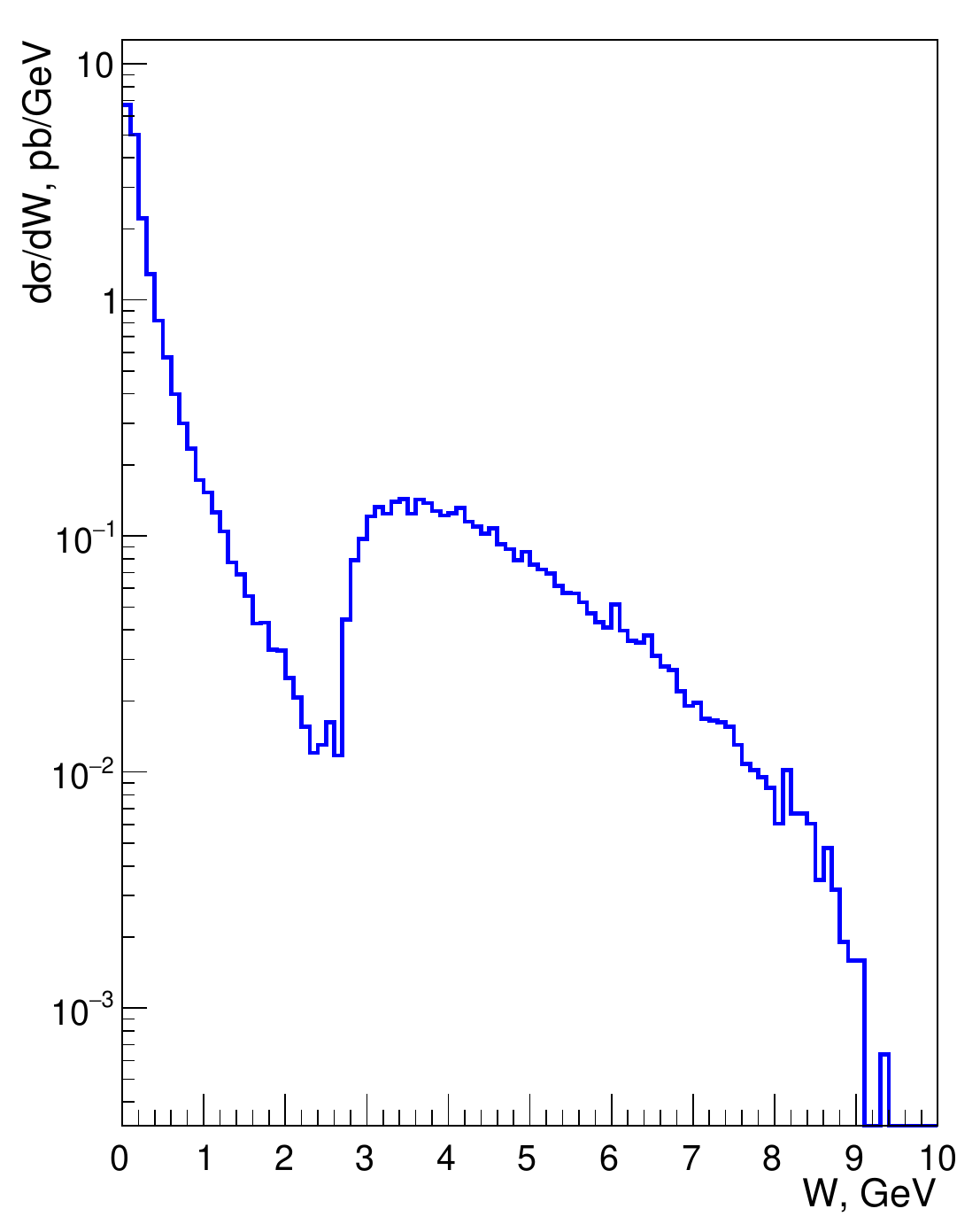} \hspace*{-0.2cm}
\includegraphics[width=4.2cm,height=4.8cm]{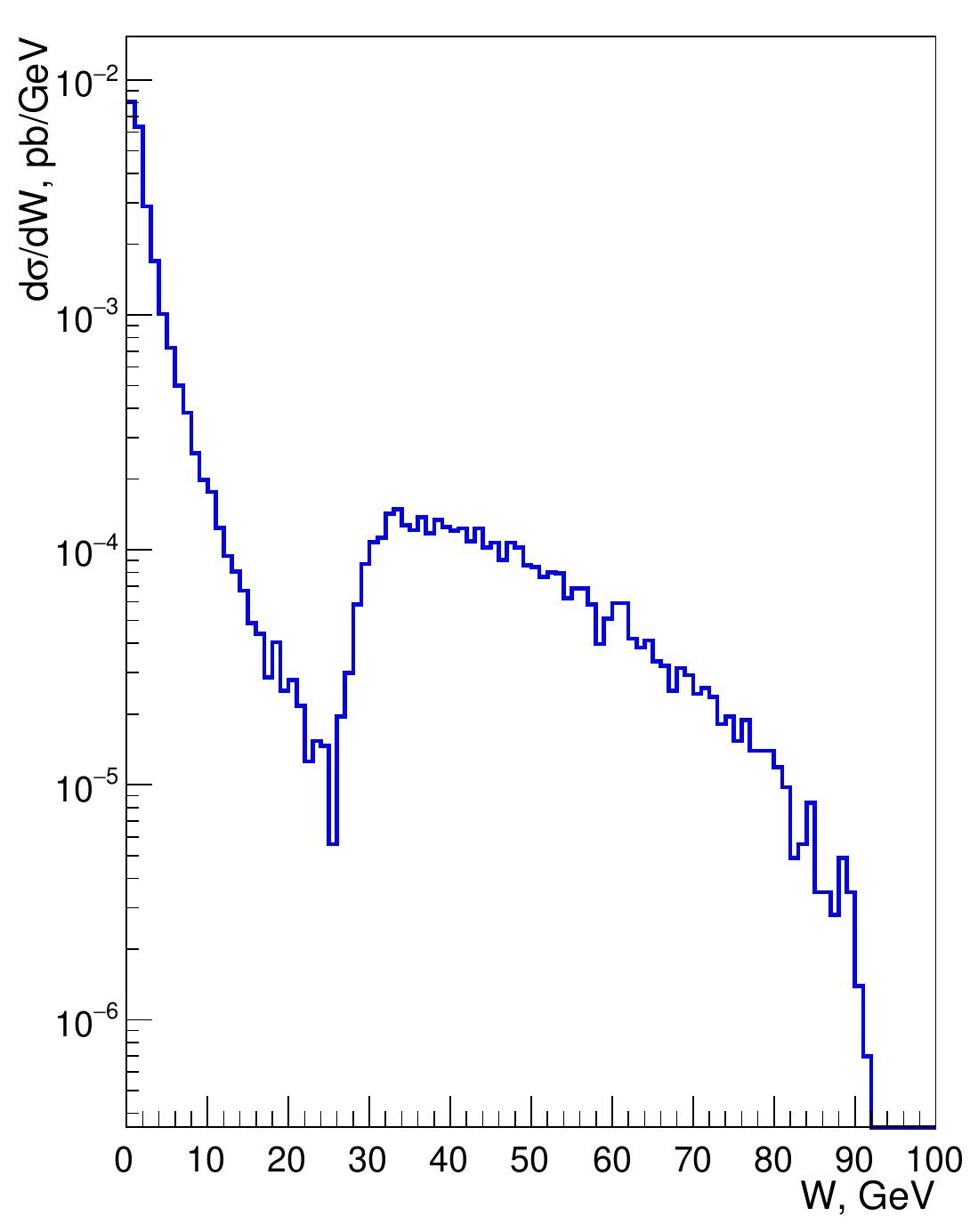}
\caption{Cross section for the $e^+e^- \to e^+e^-\gamma\gamma$ process with only angular cuts applied (see text for details).}
\label{dsdw-eegg}
\end{figure}

As before, the selection criteria required the two photons to have polar angles in the range $30^{\circ} < \theta < 150^{\circ}$, while the final-state electron and positron were restricted to small angles, $\theta < 10^{\circ}$ or $\theta > 170^{\circ}$. The cross sections with only these angular cuts applied are shown in Fig.~\ref{dsdw-eegg}.
\begin{figure}[htbp]
\centering
\includegraphics[width=4.2cm,height=4.8cm]{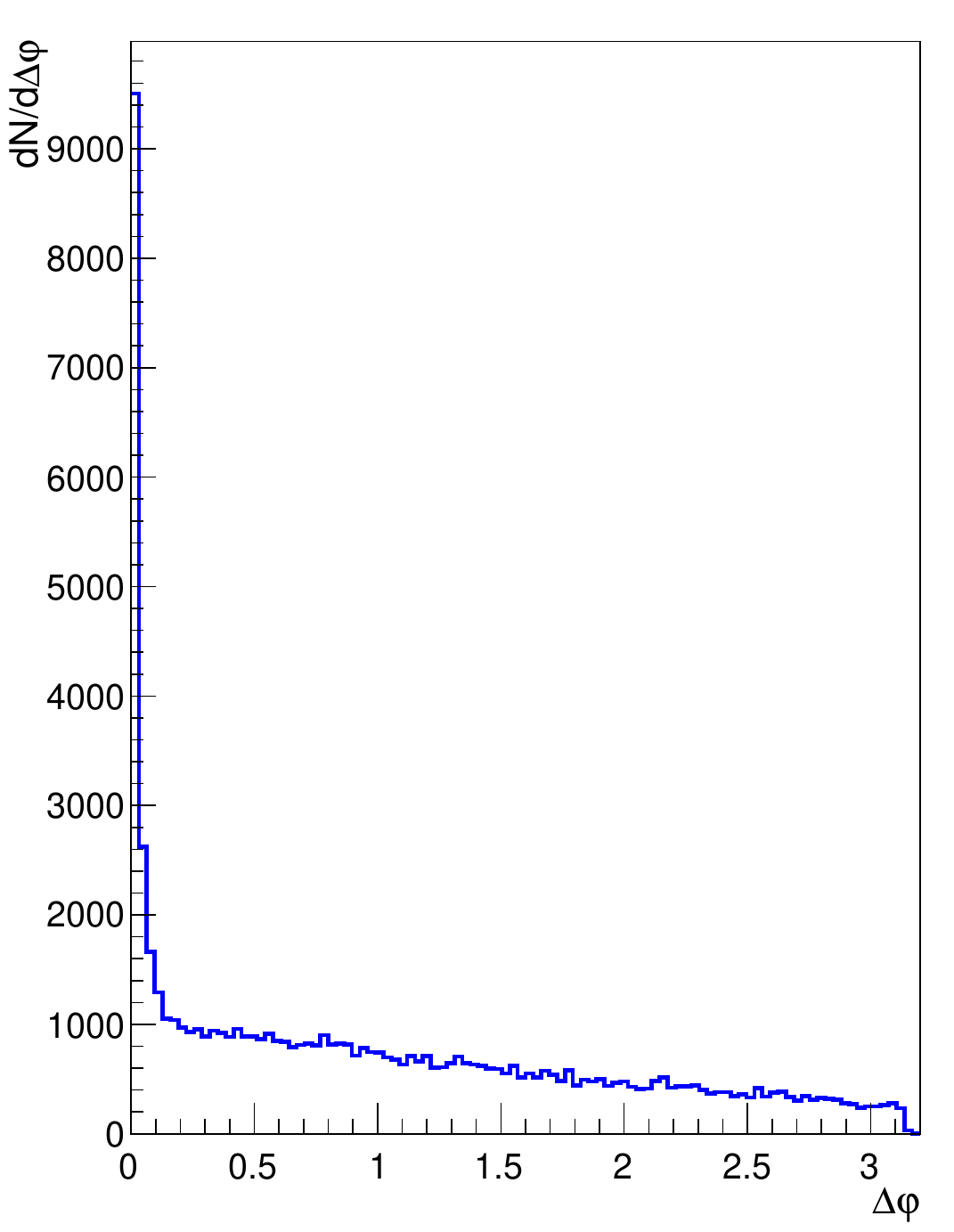} \hspace*{-0.2cm}
\includegraphics[width=4.2cm,height=4.8cm]{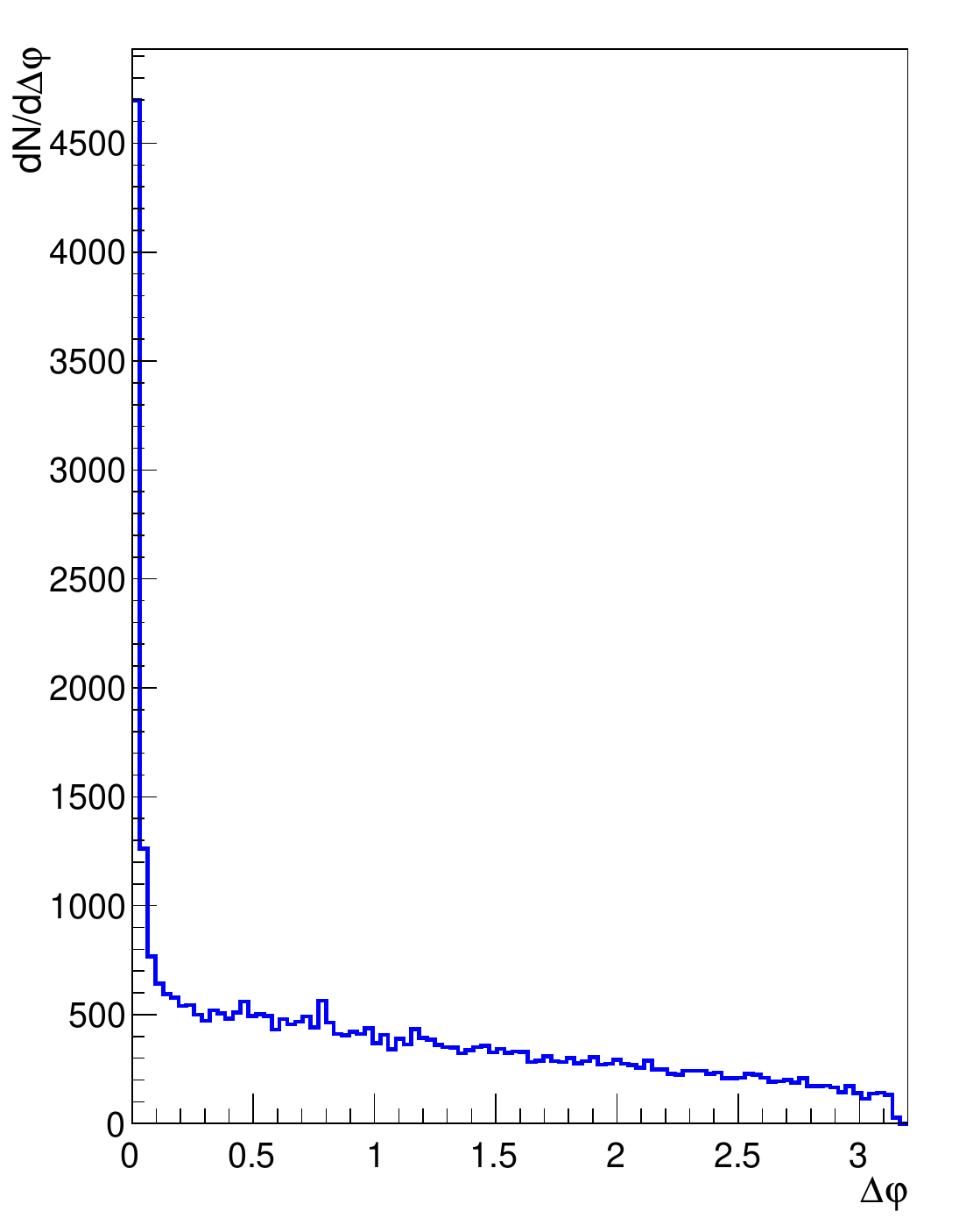}
\caption{Acoplanarity angle distribution between photons in the $e^+e^- \to e^+e^-\gamma\gamma$ process with only angular cuts applied (see text for details).}
\label{df0-eegg}
\end{figure}

The distribution of these events over the acoplanarity angle is shown in Fig.~\ref{df0-eegg}. Unexpectedly, we observe photon pairs with small acoplanarity angles. The origin of these events is as follows: it is a process of $e^+e^-$ annihilation into two photons, where one of the initial-state electrons "emits" a virtual $e^+e^-$ pair in the forward direction, while the other colliding electron (or positron) retains its beam energy $E_0$.
By using the measured energies and angles of the detected photons, we can reconstruct the energies of the colliding particles and exclude such events. We applied a cutoff $E_e < 0.9E_0$, which caused the events with small $\Delta \phi$ to disappear, along with the bumps at large $W$ in the invariant mass distributions shown in Fig.~\ref{dsdw-eegg}.

\begin{figure}[htbp]
\centering
\includegraphics[width=4.4cm]{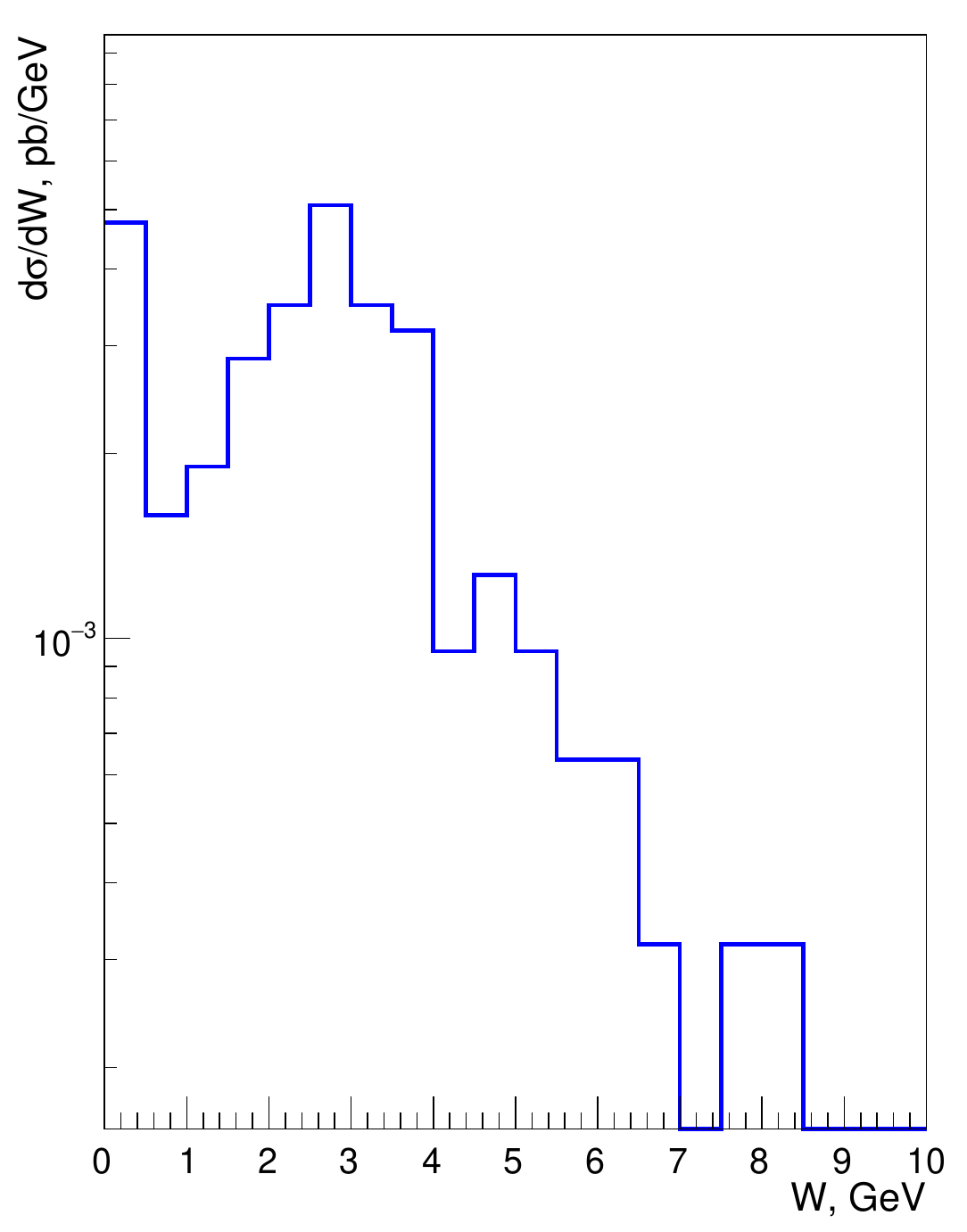}\hspace*{-0.2cm}
\includegraphics[width=4.4cm]{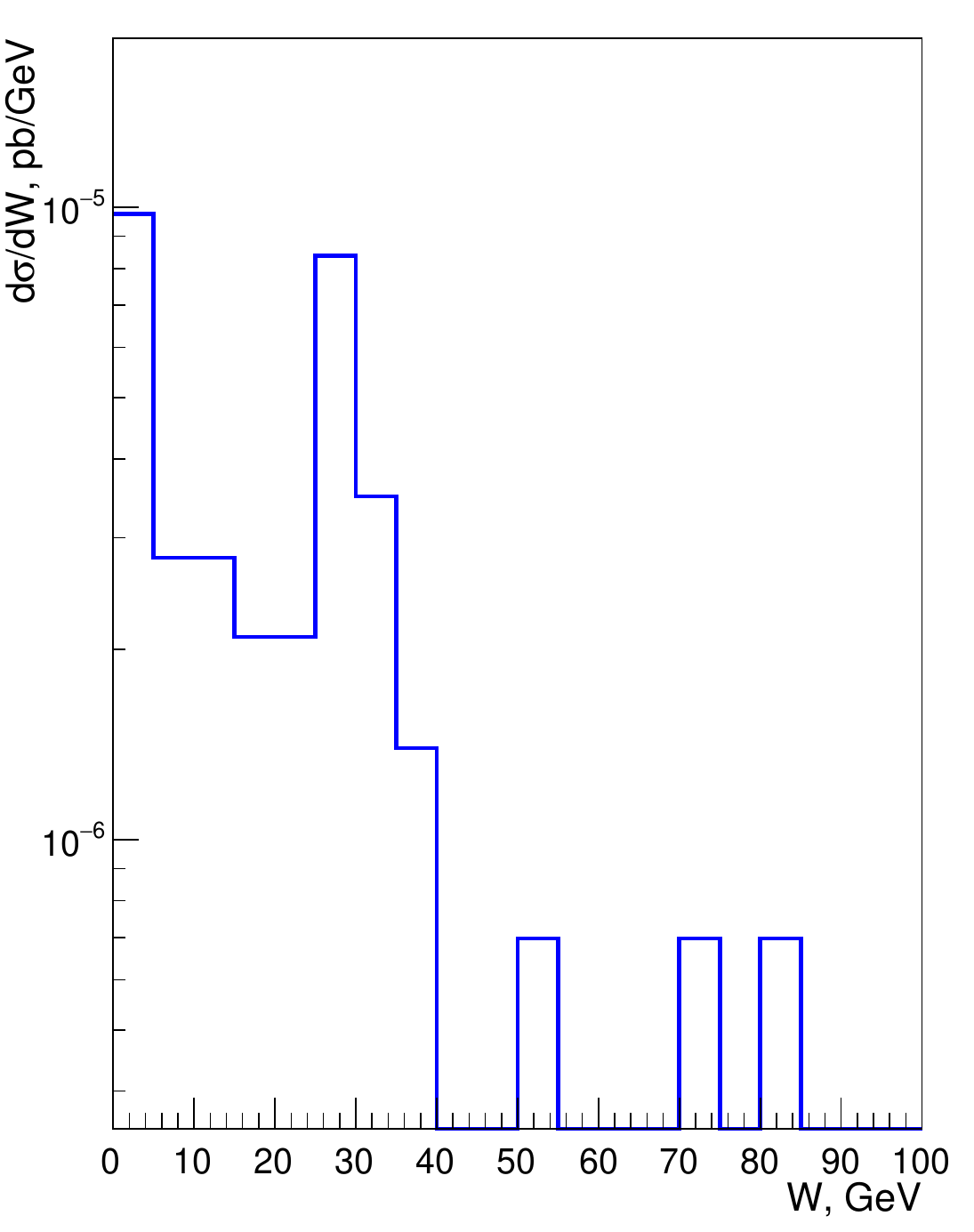}
\caption{Cross section for the $e^+e^- \to e^+e^-\gamma\gamma$ process with the cuts $E_e < 0.9E_0$, $|p_{t,1}-p_{t,2}|/(p_{t,1}+p_{t,2}) < 0.05$, and $|\Delta \phi| < 0.1$ (see text for details).}
\label{dsdw-eegg-f}
\end{figure}

The cross section with additional cuts on the relative transverse momentum difference, $|p_{t,1}-p_{t,2}|/(p_{t,1}+p_{t,2}) < 0.05$, and the acoplanarity angle, $|\Delta \phi| < 0.1$, is shown in Fig.~\ref{dsdw-eegg-f}. The distribution of these events over the acoplanarity angle is flat at small values (the corresponding figure is not shown); therefore, for $|\Delta \phi| < 0.01$, the cross section is further reduced by a factor of ten. These cross sections will be compared with the signal and other background processes in the following section.

\section{Results, discussion}
In this section, we present the combined cross-section results for all the processes discussed above.

\begin{figure*}[htbp]
\centering
\includegraphics[width=13cm]{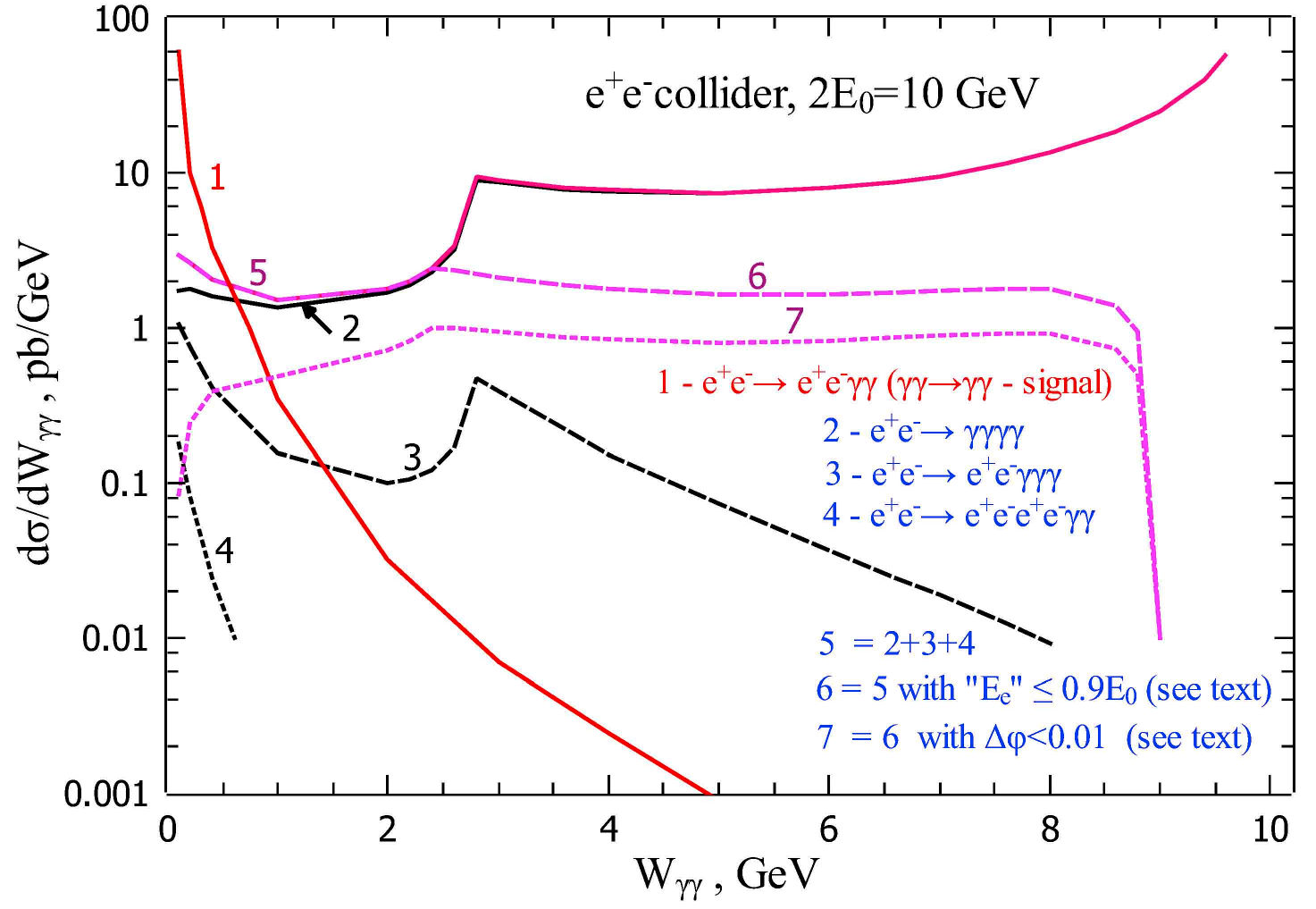}
\includegraphics[width=13cm]{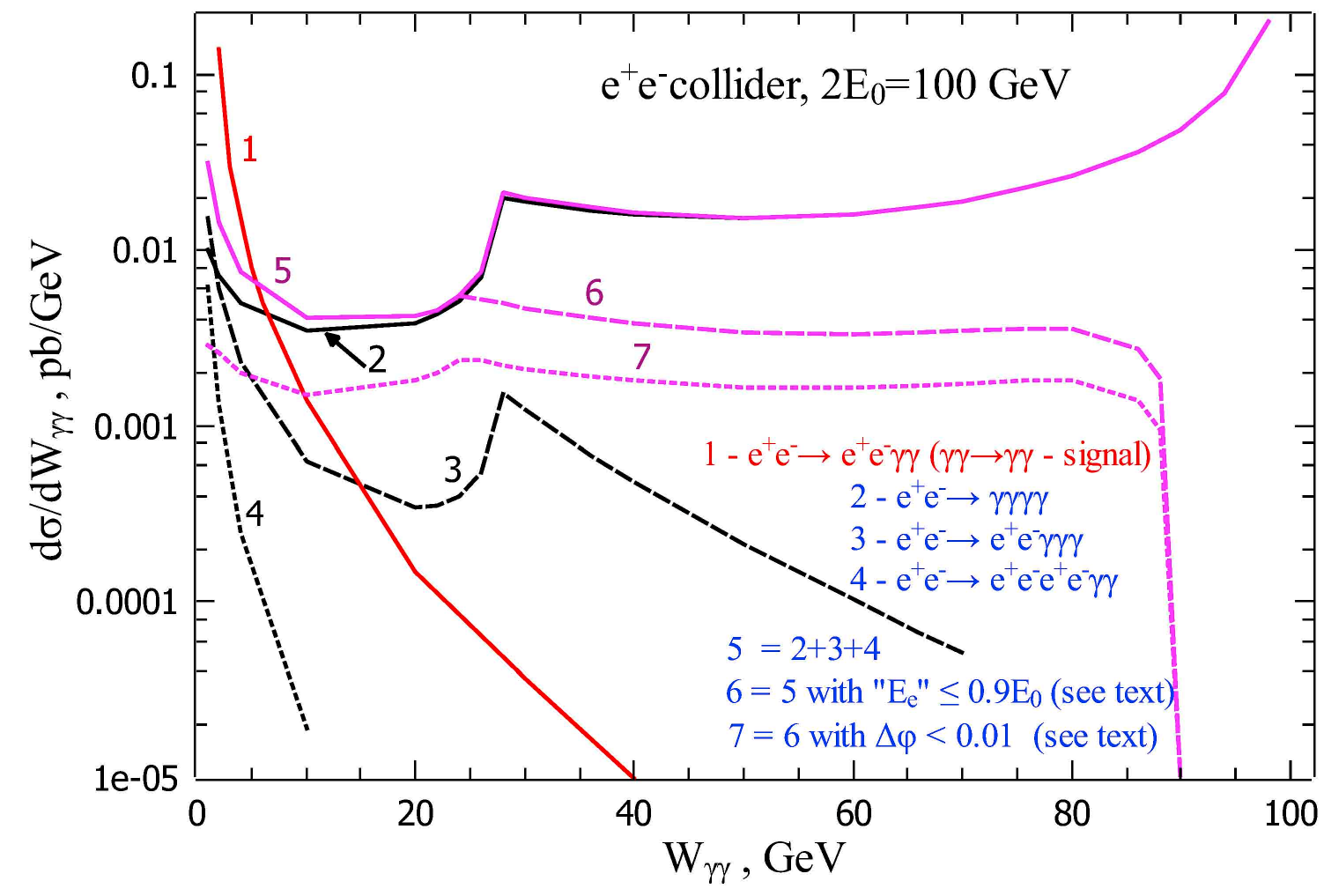}
\caption{ Cross sections of the signal (light by light scattering) and background processes as functions of the invariant mass $W_{\gamma \gamma }$ for the center-of-mass energies $2E_0=10$ GeV (upper plot) and $2E_0=100$ GeV (bottom plot). Curves 1 are the observable cross section of the $\gamma\gamma \to \gamma\gamma$ signal process at an $e^{+}e^{-}$ collider for the angular range $30^{\circ} < \theta < 150^{\circ}$ and $\Delta\phi < 0.01$. Curves 2, 3, and 4 show the cross sections for the background processes $e^+e^- \to \gamma\gamma\gamma\gamma$ (Section \ref{3}), $e^{\pm}e^- \to e^{\pm}e^-\gamma\gamma\gamma$ (Section \ref{4}), and $e^+e^- \to e^+e^-e^+e^-\gamma\gamma$ (Section \ref{5}), respectively. Curves 5 are the sums of curves 2, 3, and 4. Curves 6 are similar to curves 5, but with an additional energy cut of $E < 0.9E_0$ on the calculated energies of the particles producing the observed $\gamma\gamma$ pair. Curves 7 are similar to curves 6, with an additional acoplanarity cut of $\Delta\phi < 0.01$. The polar angular range for all background curves is identical to that of the signal, and curves 7 represent the final background level after all selection cuts.   }
\label{s-eegg}
\end{figure*}

Figure \ref{s-eegg} shows the observable cross sections of the signal (light-by-light scattering) and background processes as functions of the invariant mass $W_{\gamma \gamma }$ for the center-of-mass energies $2E_0=10$ GeV (upper plot) and $2E_0=100$ GeV (bottom plot). Curve 1 represents the cross section of the \(\gamma\gamma \to \gamma\gamma\) signal process at an $e^+e^-$ collider for the angular range \(30^{\circ} < \theta < 150^{\circ}\) and \(\Delta\phi < 0.01\). This cross section was obtained analytically using the method described in Section 2, where the angular acceptance of the detector was approximated by a rapidity cut $|\eta_{\rm max}| < 1.31$ .

Curves 2 in Fig.~\ref{s-eegg} show the cross sections for the background process $e^+e^- \to \gamma\gamma\gamma\gamma$ (Section \ref{3}). The calculations were performed under the following conditions: a maximum rapidity $\eta_{\rm max} = 1.31$ and a maximum acoplanarity angle $\Delta \phi = 0.1$. The characteristic shape of these curves is due to the fact that for $z = W_{\gamma\gamma}/2E_0 < 0.27$, the luminosity is limited by the rapidity threshold. Additionally, the cross section exhibits a logarithmic divergence as the photon energy approaches $E_{0}$ (manifested as an increase in the cross section when $W_{\gamma \gamma }$ approaches $2E_{0}$).

Curves 3 show the cross sections for the background process $e^{\pm}e^- \to e^{\pm}e^-\gamma\gamma\gamma$ (Section \ref{4}).  This background clearly exceeds the signal at approximately $W_{\gamma \gamma } > 2$~GeV for $2E_0 = 10$~GeV and $W_{\gamma \gamma } > 20$~GeV for $2E_0 = 100$~GeV. Notably, this behavior holds true for both $e^+e^-$ and $e^-e^-$ colliders.

Curves 4 show the cross sections for the background process $e^+e^- \to e^+e^-e^+e^-\gamma\gamma$ (Section \ref{5}). This background contribution is significantly smaller than the signal.

 Curves 5 represent the sum of curves 2, 3, and 4. Curves 6 are similar to curves 5, but with an additional energy cut of $E < 0.9E_0$ imposed on the energies of the particles producing the observed $\gamma\gamma$ pair. This energy cut significantly reduces the backgrounds at high invariant masses and makes the distribution smoother. Curves 7 are similar to curves 6, but includes an additional acoplanarity cut of \(\Delta\phi < 0.01\). The polar angle range for all background curves is identical to that of the signal, and curves 7 represent the final background level after all selection cuts.

 Thus, the background process $e^+e^- \to \gamma\gamma\gamma\gamma$, which occurs only at $e^+e^-$ colliders and is absent in $e^-e^-$ collisions, is dominant. Even after all cuts are applied, it exceeds the light-by-light scattering signal at $W_{\gamma \gamma } > 1$~GeV for $2E_0 = 10$~GeV and at $W_{\gamma \gamma } > 10$~GeV for $2E_0 = 100$~GeV.

The cross section for the background process $e^+e^- \to e^+e^- \gamma\gamma$, obtained using CompHEP (Section \ref{6}), is not included in the summary plot (Fig.~\ref{s-eegg}); instead, it is presented separately in Fig.~\ref{dsdw-eegg-f} for $\Delta\phi < 0.1$. For this process, the acoplanarity angle distribution of events is flat at small values; therefore, applying a tighter cut of $\Delta \phi < 0.01$ further reduces the cross section by a factor of ten. As a result, the final cross section for this process is three orders of magnitude smaller than those of the primary background processes represented by curves 7 in Fig.~\ref{s-eegg}.

In this paper, we have investigated the feasibility of studying light-by-light scattering at $e^+e^-$/$e^-e^-$ colliders in a configuration where only two scattered photons with a small total transverse momentum are registered in the central detector. We identified background QED processes that produce events virtually identical to the signal but with significantly larger cross-sections. Here, we present only the signal-to-background ratios under specific kinematic cuts optimized for background suppression, without conducting a full analysis to determine the ultimate measurement precision. However, even if the signal and background events appear identical, the processes can be statistically disentangled given sufficient luminosity by exploiting differences in their polar angle distributions. Furthermore, the QED background can be precisely subtracted, provided that high-accuracy event generators are utilized. A comprehensive analysis incorporating detector simulations and systematic uncertainties constitutes a separate task that requires a dedicated study.

Naturally, the question arises whether the scattered electrons (see Fig.~\ref{ee-box}) can be tagged directly. Indeed, implementing such tagging would suppress virtually all the backgrounds discussed in this work. Such systems are already operational; for instance, the KEDR detector at the VEPP-4M collider in Novosibirsk is equipped with a scattered electron tagging system designed for two-photon physics~\cite{KEDR}. It registers electrons with relative energy losses between 0.02 and 0.6, achieving an energy resolution of 0.1--0.3 \%. While this perfectly matches the physical requirements, such a system cannot operate at ultra-high-luminosity colliders due to the overwhelming background from luminosity bremsstrahlung. For instance, at FCC-ee running at the \(Z\)-pole, approximately 6,000 bremsstrahlung events with an electron energy loss exceeding 1\% occur per single bunch crossing (compared to about 70 events for SuperKEKB), alongside roughly 3,000 events with a loss over 7\%. Under these conditions, isolating the specific signal electrons originating from light-by-light scattering becomes practically unfeasible.

\section{Conclusion}

We have analyzed the background QED processes that  challenge the measurements of elastic light-by-light scattering at $e^+e^-$ and $e^-e^-$ colliders. The primary background arises from $e^{+}e^{-}$ annihilation into a photon pair after the initial electron and positron emit hard ISR photons at small angles. As shown in Fig.~\ref{s-eegg}, the considered QED backgrounds exceed the $\gamma\gamma \to \gamma\gamma$ signal at invariant masses $W_{\gamma \gamma } > 1$~GeV for $2E_0 = 10$~GeV (SuperKEKB) and $W_{\gamma \gamma } > 10$~GeV for $2E_0 = 100$~GeV (FCC-ee, CEPC) and become highly dominant as $W$ increases.

The situation is more favorable at $e^-e^-$ colliders, where this dominant annihilation channel is absent. Instead, the primary background originates from the annihilation of an electron (after ISR emission) with a virtual positron from the cloud of the counter-propagating electron. This background surpasses the light-by-light scattering signal at $W_{\gamma \gamma } > 2$~GeV for $2E_0 = 10$~GeV and $W_{\gamma \gamma } > 20$~GeV for $2E_0 = 100$~GeV.

Consequently, investigating $\gamma\gamma \to \gamma\gamma$ scattering at high invariant masses poses a significant challenge for both $e^+e^-$ and $e^-e^-$ colliders. Notably, these background processes are absent in ultra-peripheral heavy-ion collisions.

Photon colliders based on laser backscattering offer a promising opportunity to study light-by-light scattering at large invariant masses. Although the background environment there is complicated by $e^+e^-$ pair production both in the conversion region and coherently by photons in the field of the opposing beam, photon colliders benefit from a higher $\gamma\gamma$ luminosity and a pronounced high-energy peak in the luminosity spectrum. This kinematic feature enables effective suppression or subtraction of all relevant backgrounds.

\section*{Acknowledgments}

  This work was supported by the Russian Science Foundation (grant number 24-22-00288).

\end{document}